# Oceanic influence on Large-Scale Atmospheric Convection during co-occurring La Niña and IOD events

Supriya Ovhal[1,2], Mujumdar M[1*], Swapna P[1], Sreenivas P[3], Sandeep N[1], M. Ravichandran[4]

[1] Centre for Climate Change Research, Indian Institute of Tropical Meteorology, Ministry of Earth Sciences (MoES), Pune 411008, India

[2] Department of Atmospheric and Space Sciences, Savitribai Phule Pune University, Pune, 411007, India

[3] University of Hyderabad, Telangana 500046, India

[4] Ministry of Earth Sciences (MoES), New Delhi, India

---

Corresponding author address:

Milind Mujumdar.

Centre for Climate Change Research, Indian Institute of Tropical Meteorology

Pune 411008, India, (https://orcid.org/0000-0001-6002-5830)

e-mail: mujum@tropmet.res.in

# ABSTRACT

The Indian Summer Monsoon Rainfall (ISMR) profoundly impacts the lives of over a billion people across the region. Historically, its extremes have been linked to the El Niño/Southern Oscillation (ENSO) and modulated by the Indian Ocean Dipole (IOD). Notably, the monsoon rainfall during June to September 2022 displayed intriguing spatial patterns: above-normal precipitation over the south peninsula and central India, normal over Northwest India, and below-normal over East and Northeast India. In 2022, La Niña conditions associated with a negative Indian Ocean Dipole (nIOD) (hereafter co-occurrence years), were prevalent over the equatorial Pacific and Indian Ocean. This study investigates the often-overlooked Oceanic subsurface contribution to large-scale atmospheric convection over the tropical Indian Ocean. By undertaking a comprehensive analysis of the observed and reanalysis datasets, we find that strong equatorial westerly wind anomalies prevailed over the equatorial Indian Ocean during co-occurring years generating eastward propagating downwelling Kelvin waves which deepens the thermocline in the eastern equatorial Indian Ocean. The Kelvin waves then propagates into Bay of Bengal and gets reflected from southern tip of India as westward propagating Rossby waves deepening the thermocline and causing low level wind convergence in the northern Arabian Sea. We identify two centres of low-level wind convergence, one over Northern Arabian Sea and another over Maritime Continent during co-occurrence years. These wind convergence centres play a pivotal role in channelling moisture towards their core, in addition, significantly deepen the oceanic thermocline, increase the Ocean heat content and thus maintaining warmer SSTs over area of conversions. Together, these interrelated factors create an environment that is conducive for enhanced convective activity over the zone of convergence, enhancing and sustaining the convective conditions. This study underscores the significant contribution of ocean dynamics in shaping large-scale atmospheric convection during the co-occurrence years. Beyond the conventional focus on SST, our findings highlight the significance of considering broader Oceanic variables in comprehending and forecasting monsoonal variability.

Keywords: ISMR, ENSO, La Niña, IOD, thermocline, Ocean heat content.

## 1. Introduction:

The complex relationship between the Ocean and the atmosphere gives rise to a coupled system in which alterations in one component can exert profound effects on the other, highlighting the intricate nature of Earth's climate system. Within this framework, the connection between Sea Surface Temperature (SST) and various atmospheric processes stands as a pivotal driver of global climate dynamics, as supported by studies (Perkins et al. 2012; Zampieri et al. 2016; Zheng and Wang 2019; Hong et al. 2021). The increasing global SST trends serve as an indicator of large climate change and primary productivity in the Oceans (Luo et al. 2012; Roxy 2015; Abish et al. 2018). An exemplary illustration of this Ocean-atmosphere interaction is the El Niño-Southern Oscillation (ENSO), a coupled phenomenon centred in the tropical Pacific. This phenomenon underscores the profound influence of sub-surface Ocean conditions on SST patterns and, consequently, atmospheric processes (Bjerknes et al. 1969; Neelin et al. 1998). While ENSO theories have traditionally focused on the Pacific Ocean context, the conceptualization of Indian Ocean tropical modes, including the Indian Ocean Dipole (IOD) (Saji et al. 1999; Behera et al. 1999)and Indian Ocean Basin Mode (IOB) (Yang et al. 2007, 2010), are rooted in similar principles of Bjerknes feedback. Significantly, robust connections have been identified between SST and circulation patterns, especially during Strong IOD years, resulting in intensified thermocline-SST coupling in the Eastern Indian Ocean (EIO) and Western Indian Ocean (WIO). This leads to high SST gradients and anomalous Walker circulation within the Indian Ocean (Deshpande et al. 2014). One of the most critical climatic phenomena influenced by these dynamics is the Indian Summer Monsoon Rainfall (ISMR), occurring from June to September (JJAS), with substantial implications for Indian agriculture and the economy (Gadgil, Gadgil S 2006).

Historically, ISMR variability has been closely tied to the El Niño/Southern Oscillation (ENSO), with La Niña often leading to surplus monsoonal rainfall in India. However, the complexity of this interaction expanded with the discovery of the IOD, revealing a new layer of interannual Ocean-atmosphere coupling. This evolving understanding paved the way for strong connections between IOD events, ENSO, and their influence on ISMR, emphasizing their role in shaping regional climate dynamics and societal well-being(Ashok et al. 2001; Krishnan and Swapna 2009). Numerous studies have explored the year-to-year variations of ISMR, revealing robust links with prominent climate drivers such as ENSO, IOD, tropical Atlantic Sea surface temperature, and Middle-Eastern surface temperature. Nonetheless, the precise nature of these connections and the underlying physical mechanisms driving their interactions remain subjects of ongoing research.

As depicted in Figure 1(a), the behaviour of the Indian Summer Monsoon Rainfall (ISMR) in 2022 during co-occurrence of La Niña and nIOD exhibited a distinct and somewhat dipole pattern, characterized by both positive and negative variations in rainfall distribution across different regions of the country, as previously noted by Behera and Ratnam 2018. Notably, Figure 1(a) illustrates that the 2022 JJAS season witnessed significant rainfall, resulting in extreme weather conditions throughout the country. To gauge the prevailing drought conditions, the Standardized Precipitation Index (SPI), which relies on precipitation data, is frequently employed. In the context of the 2022 monsoon season (JJAS), Figure 1(b) provides SPI values, revealing dry conditions with negative rainfall anomalies observed in various parts of northern India, the northeast, and the Indo-Gangetic plain. Conversely, notably positive rainfall anomalies are evident over the western regions of the country (source: *https://mausam.imd.gov.in/imd_latest/contents/ar2022.pdf*). The co-occurrence of La Niña and IOD conditions and their combined influence on monsoonal disruptions over the India-Pakistan region during specific years, such as 2010 and 2022, remains a central focus. In this context, the role of large-scale circulation anomalies induced by La Niña in conjunction with unusual summer monsoon rainfall activity in sub-tropical South Asia and the tropical Indo-Pacific sector in 2022 emerges as a key research area. The interplay of circum-global teleconnection patterns, their interaction with summer monsoonal winds, and their contribution to enhanced rainfall anomalies over the region add to the complexity. Within this intricate web of Ocean-atmosphere interactions lies the question of how La Niña and IOD conditions during the summer monsoons of co-occurring years like, 2010 and 2022 influenced the Indian monsoon. Did Oceanic subsurface feedback play a role in these years? Did the La Niña-IOD relationship with Oceanic subsurface parameters significantly shape the large-scale circulation patterns sustaining monsoon disruptions across the India-Pakistan region during these years? Furthermore, a comprehensive analysis spanning from 1974 to 2022 will shed light on the evolution of La Niña-induced monsoon teleconnections over nearly five decades. Through these investigations, this research aims to deepen the understanding of the Ocean-atmosphere coupling phenomenon and its far-reaching implications for one of India's most pivotal climatic phenomena, the ISMR.

The following sections provide a comprehensive breakdown of this study's content:

In the forthcoming section, we will delve into the particulars of both the observed and model data sets, alongside the methodology employed in this research. Section 3 will spotlight the results obtained from diagnostic analysis, obtained from observed data products. These results aim to elucidate the intricate interplay between La Niña and nIOD, Oceanic subsurface parameters, large-scale circulation anomalies, and their dynamic connections with a typical summer monsoon rainfall activity across the Northern AS, north western Indian region, and adjoining Pakistan. This investigation will encompass the years 2022 and 2010. To gain a comprehensive understanding of how Ocean-atmosphere interactions have evolved over the 48 years, an extended analysis spanning the period from 1974 to 2022 was conducted. This Section will also present the outcomes derived from the model simulations and data, shedding light on their relevance to the study. The final section 4 will encompass a detailed discussion of the findings and conclude with a summary of the research's key insights.

## 2. Data and methodology:

### 2.1. Data

#### 2.1.1. Observed and model data

In this research study, a comprehensive range of datasets covering the summer monsoon season (JJAS) has been utilized, incorporating critical meteorological and Oceanographic information. These datasets are sourced from reputable institutions and provide essential details as follows: The rainfall gridded data is available at IMD database *https://imdpune.gov.in/Clim_Pred_LRF_New/Grided_Data_Download.html*. The Indian Meteorological Department (IMD). This dataset is well documented in the work by (Pai et al. 2015). To represent atmospheric circulation at standard pressure levels and mean sea level pressure, the study employs the National Center for Environmental Prediction and National Center for Atmospheric Research (NCEP/NCAR) reanalysis datasets created by Kistler et al.. Outgoing Longwave Radiation (OLR), serving as a proxy for convection, is sourced from the National Oceanic and Atmospheric Administration (NOAA). These data are interpolated and possess a spatial resolution of 2.5° × 2.5°. The dataset covers the timeframe from 1979 to 2022 and can be accessed at *http://www.cdc.noaa.gov*. Monthly SST data is derived from the Hadley Centre Global Sea Surface Temperature (HadISST) dataset developed by Rayner et al. 2003.The European

Centre for Medium-Range Weather Forecasts (ECMWF) provides the ERA5 dataset, as documented by Hersbach et al. 2020. These datasets are accessible at a grid resolution of 1° × 1° and encompass the period from 1974 to 2022. This dataset is instrumental in providing information regarding atmospheric and Oceanic signatures. Additionally, the study has incorporated data from the Coupled Model Intercomparison Project Phase 6 (CMIP6). Specifically, focus has been directed toward the picontrol dataset of the Indian Institute of Tropical Meteorology Earth System Model (IITM-ESM) as described by Swapna et al. 2018. In picontrol simulations, models are run without anthropogenic forcing, providing a baseline climate state unaffected by human activities. These datasets collectively form the foundation of the research, enabling a comprehensive exploration of monsoon dynamics and their interactions with various climate phenomena over a substantial temporal span.

### 2.2. Methodology

In this analysis, we conduct a thorough examination of both observed and model datasets on a monthly time scale. To ensure comparability, we apply a detrending process by removing linear trends from the data at each location across the entire observed data record. Additionally, we standardize all datasets, whether observed or model, to a uniform resolution of 1°×1° throughout the course of this study. Two key standard indices are established for our analysis: First, the Nino 3.4 Index, which is derived by calculating monthly SST anomalies and averaging them over the region spanning from 5° S to 5° N latitude and 170° W to 120° W longitude. Second, we identify nIOD Years by referring to information from the Australian Meteorological Office website. To pinpoint observed La Niña years, we utilize the Oceanic Niño Index (ONI), specifically identifying La Niña years when the ONI, representing a 3-month running mean of SST anomalies in the Nino 3.4 region, negative values for five consecutive seasons, covering the period from June to August through October to December. Furthermore, we ensure that identified La Niña events commence during the boreal summer (JJAS) and reach their peak during the following boreal winter. For the classification of strong La Niña years, the ONI must exceed 1 standard deviation, and a similar criterion is applied to identify strong nIOD years, specifically for the years 2010, 2022, and the model composite. Conversely, ONI and nIOD index values less than 1 standard deviation are considered for observed composite years. To rigorously evaluate the significance of our findings in this study, we employ a two-tailed Student's t-test for both the observed and model datasets. This statistical approach ensures a robust assessment of our results.

### 2.2.1. Ocean heat content

Here, OHC in $J/m^2$ for upper 700m is defined as

$$OHC = \rho_0 Cp \int_{d1}^{d2} T(z)dz \qquad (1)$$

Where, $Cp = 4185\ JKg^{-1}K^{-1}$ is the specific heat capacity of the sea water and $\rho_0 = 1025$ kg $m^{-3}$ is the reference seawater density, T(z) is the vertical profile of potential temperature and *d1*, *d2* are the corresponding reference depths.

### 2.2.2. Meridional Heat Transport

The meridional heat transport (MHT) is computed as follows:

$$MHT = \rho_0 Cp \int_{d1}^{d2} v(z)T(z)dz \qquad (2)$$

Where, $\rho_0 = 1025$ kg $m^{-3}$ is the density of water, $Cp = 4185\ JKg^{-1}K^{-1}$ the specific heat capacity of water, v is the meridional ocean velocity and T the temperature of the water column.

## 3. Results

### 3.1. Observed changes in large scale Atmospheric features during strong co-occurrence year:

This section provides an overview of the key findings and implications of the analysis for the year 2022, which is characterized by a strong La Niña and negative Indian Ocean Dipole (nIOD) according to the study. The primary objective is to summarize essential aspects of the relationship between Sea Surface Temperatures (SST) and Outgoing Longwave Radiation (OLR) to help understand the core results of the research. Figure 2a reveals a significant cooling anomaly in SST over the eastern tropical Pacific Ocean, corresponding to the 2022 La Niña event. Simultaneously, an internal variability pattern emerges with lower SST in the western equatorial Indian Ocean (WEIO) and elevated temperatures in the Eastern Equatorial Indian Ocean (EEIO), emphasizing the negative phase of the IOD. The supporting information, particularly in Figure S2, discusses the presence of similar strong La Niña and negative Indian Ocean Dipole (nIOD) conditions in the 2010 JJAS season, resembling those observed during the 2022 JJAS season. This comparison helps establish patterns and connections between these climatic events across different time periods. Figure 2b illustrates anomalous OLR values, reflecting convective and non-convective cloud conditions. Comparing Figures 2a and 2b elucidates the spatial relationship between SST and OLR. This distribution aligns with La Niña and nIOD conditions, marked by enhanced convection in the equatorial west Pacific, northwest IO, EEIO, and adjacent Indonesian region. The distribution of OLR anomalies associated with nIOD and La Niña events exhibits a zonal orientation, affecting rainfall in India and Pakistan (Behera and Ratnam 2018). Notably, the substantial westward shift of the Western Pacific Subtropical High (WPSH) in 2022 influences larger monsoonal circulation and rainfall activity over Pakistan and northwest India, corroborating findings from Ghassabi et al. 2023 for 2010. The change in WPSH positioning becomes readily evident when examining the contour map of Sea Level Pressure (SLP) anomalies in Figure 2(a). A remarkable similarity in the behaviour of OLR and SLP during the 2022 JJAS season with that of the 2010 JJAS season is documented in the supporting information, specifically in Figure S2.

### 3.2. The Influence of Ocean Dynamics on Atmospheric convection:

The transport of heat in the Indian Ocean is influenced by various climatic phenomena, including El Niño, La Niña, and the Indian Ocean Dipole (IOD). These events lead to changes in Sea surface temperatures (SSTs) and atmospheric circulation patterns, which, in turn, impact heat distribution in the Indian Ocean. The temperature gradient in ocean creates an east-west temperature gradient in the troposphere above the Indian Ocean, with areas over warmer waters experiencing increased heating. This enhanced heating can influence atmospheric convection patterns and the strength of zonal winds, affecting phenomena like the activity of Kelvin waves. These alterations in SST associated with the IOD have a significant impact on regional climate patterns, including rainfall, wind patterns, and the propagation of oceanic Kelvin waves. One key focus of a study by (Rao et al. 2010) was to examine the observed variability of Kelvin waves and their propagation in the equatorial waveguide of the Indian Ocean, as well as in the coastal waveguides of the Bay of Bengal and the south-eastern Arabian Sea, on seasonal to interannual time scales. The interannual variability of the downwelling Kelvin wave is largely driven by the strength of zonal winds over the equator, which, in turn, is influenced by east-west gradients of heat sources in the troposphere. Modelling studies by Potemra et al. 1991; Yu et al. 1991; Shankar et al. 2002) have shown the significant role of surface winds over the Eastern Indian Ocean in modulating circulation patterns in the northern Indian Ocean through remote effects. In Figure 3(a), a clear signature of Kelvin wave propagation eastward is observed during the co-occurrence of La Niña and negative IOD. Zonal wind patterns are also represented in this figure, with solid lines denoting positive (westerlies) contours and dotted lines indicating negative (easterlies) contour values. This shows strong westerly winds with positive contours over the equator, which leads to more pronounced intraseasonal moist convection signals and, consequently, stronger propagation of the downwelling Kelvin wave. In Figure 3(b), reveals the spatial distribution of Sea Level Anomalies (SLA) during the JJAS season for the co-occurrence year. Notably, the figure 3(b) exhibits distinct patterns where Kelvin waves make a prominent impact as they approach the Sumatra coast. These waves subsequently split into two directions, with northward and southward propagation of coastally trapped waves becoming evident. The northern branches propagate over varying distances along the coastal waveguide of the Bay of Bengal. These Kelvin waves also trigger Rossby waves from the eastern rim of the Bay of Bengal, which propagate westward in the northern Eastern Indian Ocean and in the Bay of Bengal. In Fig 3(b), a clear propagation of downwelling Rossby waves is observed over the southern tip of India and further westward into the Arabian Sea. This feature is particularly noticeable during events when La Niña co-occurs with nIOD. The reflected downwelling Rossby wave carries warm water and contributes to increasing the ocean heat content over the Arabian Sea. Changes in thermocline depth can be directly influenced by local winds and by remote effects through the propagation of baroclinic waves in the ocean (Schott et al. 2009) . A warm ocean surface can trigger organized moist convection under favourable meteorological conditions, generating heat in the troposphere. In Fig 1(c) and (d), during pure La Niña events, the propagation is weakened, resulting in wind contours with negative values and weaker westerlies. In Supplementary Figure S3, during events when El Niño co-occurs with a positive Indian Ocean Dipole (pIOD), Kelvin waves propagate from the west to east. However, no significant downwelling Rossby wave patterns are evident during these events.

During pure El Niño events, the equatorial waveguide appears weakened, and the amplitude of the waves diminishes due to decreasing wind contour values. The interplay of winds, Kelvin waves, and Rossby waves has significant implications for the heat distribution and circulation patterns in the Indian Ocean.

### 3.3. Observed Atmospheric patterns during strong co-occurring years:

The above finding further underscores the consistency in atmospheric conditions and convection patterns during the strong co-occurring years, emphasizing the recurring nature of these phenomena. Figure 4a depicts a distinctive anomalous zonal asymmetric circulation pattern during the JJAS period of 2022 over the Indian region. The convergence zone over the AS and EEIO plays a pivotal role in stimulating convection and intense rainfall, driven by complex atmospheric factors. The warm-moist southerly flow from the AS, evident in Figure 4c, is affected by La Niña-induced easterly anomalies over the northern Indian subcontinent (Huang-Hsiung Hsu et al.; Ghassabi et al. 2023). This, along with divergence over China, fosters convergence over the Northern AS. The Maritime continent's convergence zone is amplified by warmer SST in the EEIO, linked to the nIOD. The latitude-pressure section of the monsoon Hadley-type circulation for JJAS 2022 (Figure 4b) emphasizes ascending motions around the 20°N latitude, closely tied to the convective activity of the South Asian Monsoon (SAM). Simultaneously, a corresponding ascending branch within a divergent zone is observed over China, emphasizing the dynamic interaction between circulation features. The conversion pattern observed during the 2022 JJAS season closely resembles that of the 2010 JJAS season. Additionally, the circulation pattern and the presence of southerly winds exhibit similarities between the 2010 JJAS and 2022 JJAS, as detailed in Figure S4 of the paper. This parallelism suggests a recurring atmospheric and Oceanic configuration between the two years, highlighting the importance of these patterns in influencing the weather and climate during these specific monsoon seasons. Previous research indicates that strong IOD years lead to robust thermocline-SST coupling in the Eastern Indian Ocean (EIO) and Western Indian Ocean (WIO). This coupling contributes significantly to intensified SST gradients, shaping the anomalous Walker circulation within the Indian Ocean (Deshpande et al. 2014). However, the precise mechanisms underlying the influence of Oceanic subsurface conditions on SST and its seasonal variations remain subjects that require further exploration. Numerous studies have examined subsurface interannual variability within the Indian Ocean and have sought to establish connections with either the IOD through research by Rao et al. 2002 and Vinayachandran et al. 2002, or the El Niño-Southern Oscillation (ENSO) as investigated by Tourre and W. B. White; Chambers et al. 1999and Xie et al. 2002. Despite the valuable insights provided by these studies, a comprehensive understanding of the intricate relationships between subsurface Ocean dynamics and SST, particularly in relation to seasonal variations, remains an ongoing research priority. Therefore, the next section of the study delves into the significance of Oceanic variables in maintaining convection over the Northern AS and the EEIO. This examination aims to shed light on the pivotal role that Oceanic factors play in supporting the atmospheric processes and convection patterns observed in these regions.

### 3.4. Possible physical mechanism responsible for sustaining atmospheric convection:

After the 1950s, the Tropical Indian Ocean (TIO) underwent significant warming, a phenomenon well-documented in studies such as Alory et al. 2007; Du and Xie 2008. Notable dynamic features of the AS during summer include the intense coastal currents along Somalia and Oman, as well as eastward/south-eastward currents across the AS (Schott and Mccreary 2001). These upper Ocean currents are primarily driven by the Ekman transport associated with south-westerly winds in the summer, which, in turn, influence the circulation and heat transport in the

Arabian Sea. The vertical current also brings nutrient-rich water to the surface, making it of significant scientific interest (Bauer et al. 1990; S. Prasanna Kumar et al. 2001; Fischer et al. 2002; Wiggert et al. 2005; de Boyer Montégut et al. 2007). Therefore, it is crucial to comprehend how variations in the ISM circulation impact the dynamics and thermodynamics of the Arabian Sea, a topic explored in studies like Schott 1983; Keen et al. 1997; Murtugudde et al. 2007. Long-term observations and coupled model experiments have revealed a rapid warming trend in the Arabian Sea, particularly during summer (Koll Roxy et al.; Roxy 2015). The weakened southward heat transport, as a consequence of a less pronounced cross-equatorial cell, leads to increased subsurface temperatures in the North Indian Ocean (NIO), as illustrated in Figure 5a, which displays OHC (shaded) and D20 (contour) for the year 2022 JJAS. The diminished southward meridional heat transport allows heat to accumulate in the NIO. Figure 5b, which represents meridional heat transport over the region spanning 45° to 75° east and from 0° to 24°N for the year 2022 JJAS, shows a rise in warm water transport from the equator towards the pole. Starting in July, warming signals become distinctly visible in the northern region, and these signals extend with depth. The meridional heat transport in the IO is predominantly shaped by the summer monsoon circulation. This transport is propelled by wind-induced Ekman transport, causing a transfer of heat across the equator (Schott and Mccreary 2001). However, a weakening of the monsoon circulation has diminished the southward Ocean heat transport, leading to a build-up of heat in the NIO (Swapna et al. 2017). Our analysis indicates that the strong southerly behaviour of summer monsoon winds has weakened the southward Ocean heat transport, leading to a decrease in the cross-equatorial cell and a consequent rise in subsurface temperatures, particularly in the AS. Similarly, in Figure 6a, we observe positive OHC and deep D20 over the EEIO. Figure 6a for the year 2010 JJAS reveals a positive OHC (shaded) and a deepening of D20 (contour) in the northern AS and the EEIO. This pattern closely resembles the findings from Figure 5a for the year 2022. Furthermore, Figure 6b exhibits a Meridional Heat Transport (MHT) signature similar to that observed in the 2022 JJAS data, indicating consistent trends across the years. Numerous studies have proposed various factors contributing to the warm subsurface conditions in the EEIO, including the Indian Throughflow (ITF)(Wyrtki 1961). The easterly winds result in the accumulation of warm water in the western Pacific. A portion of this warm water is transported into the southern Tropical Indian Ocean (STIO) through the Indonesian islands, primarily in the form of the Indonesian Throughflow (ITF). This influx of warm water significantly boosts the upper Ocean heat content in the STIO, a phenomenon well-documented in studies by Zhang and Han; Lee et al. 2015; Liu et al. 2016; Li et al. 2018, 2020 and it has been extensively studied, as highlighted by Wijffels and Meyers 2004. However, our study, as depicted in Figure 4a, reveals the presence of a conversion zone in the EEIO, similar to the northern Arabian Sea region. This conversion zone is associated with anomalous warming in the subsurface of the EEIO. Rao and Behera 2005)explored the influence of subsurface conditions on SST variability in both the northern and southern TIO and identified two distinct zones of subsurface influence on SST. In our study, we find that the northern AS and EEIO exhibit positive SST values and warm subsurface temperatures. Consequently, the warm subsurface conditions contribute to the persistence of warm SST in these regions and create favourable conditions for strong convection in the northern Arabian Sea and EEIO.

**3.5. Long term analysis of observational records:**

As a result of our investigation into the anomalous convection and circulation patterns experienced during the summer monsoons of 2010 and 2022 within the Asian monsoon sector, we decided to extend our analysis over a more extensive timeframe, spanning from 1974 to 2022. Our primary objective in conducting this extended analysis was to gain a comprehensive understanding of historical instances of atypical monsoon responses occurring during the co-occurrence years. Figure 7(a) provides a spatial distribution of SST values for the composite years of 1974, 1998, and 2016, coinciding with the occurrence of both La Niña and the nIOD. When comparing the spatial distribution of SST for the years 2010 and 2022 with these composite years, we observed several notable distinctions. Particularly striking is the observation that the EEIO exhibited higher SSTs during both 2010 and 2022, aligning with the composite years characterized by the presence of La Niña and nIOD. However, this warming pattern was not evident over the Northern AS. Furthermore, when examining convective activity, indicated by the observation of low OLR values, it became apparent that such activity was not notably present over the Northern AS. Indeed, the presence of convective activity over the EEIO during these periods is noteworthy and supported by various studies that have elucidated the warming trends and explained the reasons for concurrent convective activity in this region (Annamalai et al. 2003; Roxy 2014; Goswami 2023). Zhou et al. 2009 previously documented a westward extension of the boreal summer WPSH since the late 1970s, attributing this phenomenon to atmospheric responses linked to observed warming in the Indian Ocean and western Pacific (IWP). Building on this foundation, Mujumdar et al. 2012drew attention to a noteworthy aspect of the 2010 summer monsoon – the pronounced westward shift of the WPSH, which was associated with La Niña conditions in the tropical Pacific. This shift in the WPSH's position was also evident during the 2022 JJAS period, as demonstrated in Figure 2(a). However, during La Niña and nIOD composite years, we observed an SLP pattern characterized by negative values over central China and the adjacent Indian region. A comparison of the SLP anomalies for 2010 JJAS (Figure S2(a)), 2022 JJAS (Figure 2(a)), and the composite years featuring La Niña and nIOD (Figure 7(a)) reveals a north eastward shift in SLP, underscoring the uneven distribution of rainfall during the years 2010, 2022, and the composite years marked by both La Niña and nIOD. In Figure S7(a) of our analysis, we identified the presence of convergence zones, one over the western part of the Arabian Sea, another was located over the Maritime continent during July and September, and additional convergence zones were observed encompassing the Indian region. In contrast, during 2010 and 2022 JJAS, we observed two convergence zones, one over the Northern AS and the other over the EEIO. These convergence zones of composite years generate updrafts over areas of convergence and downdrafts over areas of divergence, as depicted in Figure S7(b). During this period, the prevailing wind direction is predominantly southern-westerly, as indicated in Figure S7(c), resulting in the transport of moisture toward the central part of India. The convergence zone over the AS and EEIO plays a pivotal role in stimulating convection and intense rainfall, driven by complex atmospheric dynamics. However, this mechanism seems to be absent in the composite years featuring both La Niña and nIOD. While we did observe OHC and a deepening of D20 during the composite years, as shown in Figure 8(a), covering a substantial portion of the IO, it's important to note that these characteristics were less pronounced over the Northern AS. Additionally, Figure 8(b) illustrates that the meridional heat transport does not demonstrate a substantial strength during these composite years. This limited heat transport results in less heat accumulation and retention over the northern part of the AS, thereby failing to sustain the convection over the Northern AS.

### 3.6. Simulated features from IITM-ESM coupled model:

In this section, we employed Pi-Control data from the IITM-ESM model to assess the robustness of our findings in comparison to observations. Our objective was to illustrate the relationship between Oceanic subsurface conditions and atmospheric convection. Remarkably, the model output closely resembles to the results obtained from observed datasets, for strong co-occurrence years. The model simulations revealed a significant cooling anomaly in SST over the eastern tropical Pacific Ocean, corresponding to the occurrence of a strong La Niña event in the model composite. Concurrently, an internal variability pattern emerged, characterized by lower SST in the WEIO and elevated temperatures in the EEIO, resulting in the negative phase of the IOD. Further support for our findings, especially in Figure S8(a), underscored the presence of strong La Niña and nIOD. Figure S8(b) shows significant negative OLR anomalies over the western part of Indian region and EEIO events resembling those observed during the 2010 and 2022 JJAS seasons. In Figure 9(a), we observed two convergence zones, one covering the AS, encompassing the central and northern AS portions, and another over the EEIO. This configuration was influenced by the warm-moist southerly flow from the AS, as evident in Figure 9(c), which was impacted by La Niña-induced easterly anomalies over the northern Indian subcontinent. This, coupled with divergence over China, contributed to convergence over the northern Arabian Sea. The Maritime continent's convergence zone was amplified by warmer SST in the EEIO, linked to the nIOD. The latitude-pressure section of the monsoonal Hadley-type circulation for JJAS of model composite years (Figure 9(b)) highlighted ascending motions around the 20°N latitude, closely associated with the convective activity of the SAM. Simultaneously, a corresponding ascending branch within a divergent zone was observed over China, underscoring the dynamic interaction between circulation features. The observed pattern of convergence during the model composite years closely resembled that of the 2010 and 2022 JJAS seasons. Additionally, from Figure 10(a) we noted a positive OHC and deepening of D20 over the EEIO compared to the Northern AS. Importantly, in Figure 10(b) the MHT remained consistently strong with depth over north of 8°N from the month of July onwards. This was aligned with the positive SST values observed over the AS and EEIO, as illustrated in Figure S9(a). Consequently, the warm subsurface conditions played a pivotal role in maintaining warm SST in these regions, creating favourable conditions for robust convection in the northern Arabian Sea and EEIO.

## 4. Discussion and Summary

This study focuses on understanding the complex interactions between Oceanic and atmospheric processes and their pivotal role in sustaining convection patterns over the Northern tropical Indian Ocean, especially over Arabian Sea (AS) and the Eastern Equatorial Indian Ocean (EEIO). This research also investigates the influence of La Niña and Indian Ocean Dipole on the Indian Summer Monsoon Rainfall (ISMR), which has significant impact over a billion people in the region. Previous studies have examined the impact of La Niña and nIOD events on the Indian subcontinent (Mujumdar et al. 2012; Behera and Ratnam 2018)and its adjoining region, Pakistan(Huang-Hsiung Hsu et al.; Ghassabi et al. 2023). Our study examines the concurrent presence of La Niña and a negative IOD (nIOD) and their impact on the sub-surface conditions and large-scale convection and ISMR variability. The outcomes reveal intriguing patterns in monsoon rainfall during these co-occurrence years. Notably, there was below-normal precipitation observed over northern, north-eastern, and Indo-Gangetic plain regions of India, while above-normal rainfall was noted in the western part of the country and adjoining Indo-Pak region, especially during strong co-occurring years of 2022 and 2010. The study emphasizes the significance of looking beyond SST and considers broader Oceanic variables, particularly subsurface dynamics, in the understanding of the role of Ocean with respect to sustaining Atmospheric convection.

Typically, La Niña conditions, characterized by cooler SSTs in the tropical Pacific, have been associated with surplus monsoonal rainfall in India. Moreover, this study brings to light the complex role of a concurrent nIOD in this interaction, which can lead to modulation in the monsoon. The intricate interplay between these two phenomena results in diverse rainfall patterns across India. Importantly, the research identifies two key regions of low-level wind convergence one over the Northern AS and the other over the Maritime Continent. Through a thorough examination of both observed and reanalysis datasets, our analysis reveals the prevalence of robust westerly wind anomalies along the equatorial Indian Ocean during concurrent years. These anomalies give rise to the formation of eastward propagating downwelling Kelvin waves, which subsequently lead to the deepening of the thermocline in the eastern equatorial Indian Ocean. These Kelvin waves further propagate into the Bay of Bengal, where they are reflected by the southern tip of India as westward propagating Rossby waves. These Rossby waves play a crucial role in deepening the thermocline and inducing low-level wind convergence in the northern Arabian Sea. These convergence zones play a pivotal role in channelling the moisture towards warmer SSTs over the area of convergence, and thus sustaining convective activity. Thus, the study focuses on the often-overlooked role of Oceanic subsurface dynamics in shaping large-scale atmospheric convection, especially during co-occurrence years. Beyond SST, factors such as the deepening of the thermocline, increased oceanic heat content and warmer SSTs in the convergence zones of the northern AS and EEIO are recognized as crucial in enhancing monsoonal convection. These wind convergence centres seem to be associated with deepened thermoclines, higher OHC, and warmer SSTs, creating a conducive environment for intensified monsoonal convection. The study goes beyond a limited timeframe, extending its analysis from 1974 to 2022, to gain insights into the La Niña and IOD induced monsoon teleconnections. Notably, the investigation highlights the role of Oceanic subsurface conditions in sustaining atmospheric convection, especially in the Northern AS and the EEIO. Model simulations using the IITM-ESM model are analysed, which align closely with observed data. The 200-year long simulation from IITM-ESM provides longer-time series to understand the role of oceanic subsurface influence on atmospheric convection during co-occurring years.

In conclusion, this research underscores the importance of incorporating Oceanic subsurface dynamics alongside SST in the comprehension and sustaining Atmospheric convective over the region of convergence during co-occurrence years of La Niña and nIOD. It offers valuable insights into the intricate Ocean-atmosphere interactions shaping the Indian monsoon, which are very valuable for advancements in monsoon forecasting and climate modelling. The study's findings significantly contribute to our understanding of one of India's most critical climatic phenomena and its broader implications for society and agriculture in the region.

**Acknowledgments**

We sincerely thank Dr. R. Krishnan, Director, Indian Institute of Tropical Meteorology (IITM, India) and Dr. J. Sanjay, executive Director, Centre for Climate Change Research (CCCR at IITM) for all the support during the research study. We also thank IITM-ESM group for making the coupled model data available and also for using the HPC resources at IITM. We thank Mr. Ajinkya Aswale for his support in graphic illustration with NCL.

**Funding**

The authors have no funding reference to declare.

**Data Availability**

Enquiries about data availability should be directed to the authors.

**Conflict of interest**

The authors have not disclosed any competing interests.

**Author Contributions**

Supriya Ovhal initiated and led this study. Milind Mujumdar, Swapna P conceived and supervised the study. Supriya Ovhal performed the analyses supported by Milind Mujumdar Swapna P and Sandeep N. Supriya Ovhal wrote the original draft, reviewed by Milind Mujumdar, Swapna P. All authors participated in discussions and contributed to review and editing the manuscript.

**References**

Abish B, Cherchi A, Ratna SB (2018) ENSO and the recent warming of the Indian Ocean. International Journal of Climatology 38:203–214. https://doi.org/10.1002/joc.5170

Alory G, Wijffels S, Meyers G (2007) Observed temperature trends in the Indian Ocean over 1960-1999 and associated mechanisms. Geophys Res Lett 34:.2. https://doi.org/10.1029/2006GL028044

Anderegg WRL, Schwalm C, Biondi F, et al (2015) Pervasive drought legacies in forest ecosystems and their implications for carbon cycle models. Science (1979) 349:528–532. https://doi.org/10.1126/science.aab1833

Annamalai H, Murtugudde R, Potemra J, et al (2003) Coupled dynamics over the Indian Ocean: Spring initiation of the Zonal Mode. Deep Sea Res 2 Top Stud Oceanogr 50:2305–2330. https://doi.org/10.1016/S0967-0645(03)00058-4

Ashok K, Guan Z, Yamagata T (2001) Impact of the Indian Ocean dipole on the relationship between the Indian monsoon rainfall and ENSO. Geophys Res Lett 28:4499–4502. https://doi.org/10.1029/2001GL013294

Bauer S, Hitchcock GL, Olson DB (1990) Influence of monsoonally-forced Ekman dynamics upon surface layer depth and plankton biomass distribution in the Arabian Deep Sea Research Part A. Oceanographic Research Papers. 38. 531–553. https://doi.org/10.1016/0198-0149(91)90062-K.

Behera SK, Krishnan R, Yamagata T (1999) Unusual ocean-atmosphere conditions in the tropical Indian Ocean during 1994. Geophys Res Lett 26:3001–3004. https://doi.org/10.1029/1999GL010434

Behera SK, Ratnam JV (2018) Quasi-asymmetric response of the Indian summer monsoon rainfall to opposite phases of the IOD. Sci Rep 8:123. https://doi.org/10.1038/s41598-017-18396-6

Bjerknes J (1969) Atmospheric teleconnections from the equatorial Pacific. Mon Weather Rev 97:163–172

Chambers DP, Tapley BD, Stewart RH (1999) Anomalous warming in the Indian Ocean coincident with El Niño. J Geophys Res Oceans 104:3035–3047. https://doi.org/10.1029/1998jc900085

de Boyer Montégut C, Mignot J, Lazar A, Cravatte S (2007) Control of salinity on the mixed layer depth in the world ocean: 1. General description. J Geophys Res 112:C06011. https://doi.org/10.1029/2006JC003953

Deshpande A, Chowdary JS, Gnanaseelan C (2014) Role of thermocline-SST coupling in the evolution of IOD events and their regional impacts. Clim Dyn 43:163–174. https://doi.org/10.1007/s00382-013-1879-5

Du Y, Xie SP (2008) Role of atmospheric adjustments in the tropical Indian Ocean warming during the 20th century in climate models. Geophys Res Lett 35(8):391–405 https://doi.org/10.1029/2008GL033631

Fischer AS, Weller RA, Rudnick DL, Eriksen CC, Lee CM, Brink KH, Fox CA, Leben RR (2002) Mesoscale eddies, coastal upwelling, and the upper-ocean heat budget in the Arabian Sea. Deep-Sea Res II 49(12):2231–2264

Friedrich A. Schott, Julian P. McCreary, (2001) The monsoon circulation of the Indian Ocean,

Gadgil S, Gadgil S (2006) The Indian monsoon, GDP and agriculture. Econ Polit Wkly 41(47):4887–4895. https://doi.org/10.2307/4418949

Ghassabi Z, Karami S, Vazifeh A, Habibi M (2023) Investigating the unprecedented summer 2022 penetration of the Indian monsoon to Iran and evaluation of global and regional model forecasts. Dynamics of Atmospheres and Oceans 101386. https://doi.org/10.1016/j.dynatmoce.2023.101386

Goswami BB (2023) Role of the eastern equatorial Indian Ocean warming in the Indian summer monsoon rainfall trend. Clim Dyn 60:427–442. https://doi.org/10.1007/s00382-022-06337-7

Hersbach H, Bell B, Berrisford P, et al (2020) The ERA5 global reanalysis. Quarterly Journal of the Royal Meteorological Society 146:1999–2049. https://doi.org/10.1002/qj.3803

Hong CC, Tseng WL, Hsu HH, et al (2021) Relative contribution of trend and interannually varying SST anomalies to the 2018 heat waves in the extratropical northern hemisphere. J Clim 34:6319–6333. https://doi.org/10.1175/JCLI-D-20-0556.1

Hsu, H.H., Hong, C.C., Huang, A.Y., Tseng, W.L., Lu, M.M, Chang, C.C., (2022) Linkage between Record Floods in Pakistan and a Severe Heatwave in China in the Boreal Summer of 2022. Preprint. Doi:10.21203/rs.3.rs-2353452/v1.

Keen TR, Kindle JC, Young DK (1997) The interaction of southwest monsoon upwelling, advection and primary production in the northwest Arabian Sea, Journal of Marine Systems, Volume 13, https://doi.org/10.1016/S0924-7963(97)00003-1

Kistler R et al (2001) The NCEP-NCAR 50-year reanalysis: monthly means CD-ROM and documentation. Bull Am Meteor Soc 82:247–267

Krishnan R, Swapna P (2009) Significant influence of the boreal summer monsoon flow on the Indian Ocean response during dipole events. J Clim 22:5611–5634. https://doi.org/10.1175/2009JCLI2176.1

Kumar SP, Madhupratap M, Kumar MD et al (2001) High biological productivity in the central Arabian Sea during the summer monsoon driven by Ekman pumping and lateral advection. Curr Sci 81:1633–1638

Lee SK, Park W, Baringer MO, et al (2015) Pacific origin of the abrupt increase in Indian Ocean heat content during the warming hiatus. Nat Geosci 8:445–449. https://doi.org/10.1038/NGEO2438

Li M, Gordon AL, Gruenburg LK, Wei J, Yang S (2020) Interannual to decadal response of the Indonesian throughflow vertical profile to Indo-Pacific forcing. Geophys Res Lett 47:11 https://doi.org/10.1029/2020GL087679

Li M, Gordon AL, Wei J, Gruenburg LK, Jiang G (2018) Multi-decadal timeseries of the Indonesian throughflow. Dyn Atmos Oceans 81:84–95 https://doi.org/10.1016/j.dynatmoce.2018.02.001

Liu W, Xie SP, Lu J (2016) Tracking Ocean heat uptake during the surface warming hiatus. Nat Commun 7:1–9. https://doi.org/10.1038/ncomms10926

Luo JJ, Sasaki W, Masumoto Y (2012) Indian Ocean warming modulates Pacific climate change. Proc Natl Acad Sci U S A 109:18701–18706. https://doi.org/10.1073/pnas.1210239109

Mujumdar M, Preethi B, Sabin TP, et al (2012) The Asian summer monsoon response to the La Niña event of 2010. Meteorological Applications 19:216–225. https://doi.org/10.1002/met.1301

Murtugudde R, Seager R, Thoppil P (2007) Arabian Sea response to monsoon variations. Palaeoceanography 22:PA4217. doi:10.1029/2007PA001467

Neelin JD, Battisti DS, Hirst AC, et al (1998) ENSO theory. J Geophys Res Oceans 103:14261–14290. https://doi.org/10.1029/97jc03424

Pai DS, Sridhar L, Badwaik MR, Rajeevan M (2015) Analysis of the daily rainfall events over India using a new long period (1901–2010) high resolution (0.25° × 0.25°) gridded rainfall data set. Clim Dyn 45:755–776. https://doi.org/10.1007/s00382-014-2307-1

Perkins SE, Alexander LV, Nairn JR (2012) Increasing frequency, intensity and duration of observed global heatwaves and warm spells. Geophys Res Lett 39:1–5. https://doi.org/10.1029/2012GL053361

Potemra JT, Luther ME, O'Brien JJ (1991) The seasonal circulation of the upper ocean in the Bay of Bengal. J Geophys Res 96:12,667–12,683. doi:10.1029/91JC01045 Progress in Oceanography, 12:0079-6611, https://doi.org/10.1016/0079-6611(83)90014-9.

Rao RR, Girish Kumar MS, Ravichandran M, Rao AR, Gopalakrishna VV, Thadathil P (2010) Interannual variability of Kelvin wave propagation in the wave guides of the equatorial Indian Ocean, the coastal Bay of Bengal and the Southeastern Arabian Sea during 1993–2006. Deep-Sea Res I 57:1–13

Rao SA, Behera SK (2005) Subsurface influence on SST in the tropical Indian Ocean: Structure and interannual variability. Dynamics of Atmospheres and Oceans 39:103–135. https://doi.org/10.1016/j.dynatmoce.2004.10.014

Rao SA, Behera SK, Masumoto Y (2002) Interannual subsurface variability in the Tropical Indian Ocean with a special emphasis on the Indian Ocean Dipole. 49:1549–1572

Rayner NA, Parker DE, Horton EB, Folland CK, Alexander LV, Rowell DP, Kent EC, Kaplan A (2003) Global analyses of sea surface temperature, sea ice, and night marine air temperature since the late nineteenth century. J Geophys Res Atm 108(D14):4407. https://doi.org/10.1029/2002jd002670

Rev Geophys 47(1)

Roxy M (2014) Sensitivity of precipitation to sea surface temperature over the tropical summer monsoon region-and its quantification. Clim Dyn 43:1159–1169. https://doi.org/10.1007/s00382-013-1881-y

Roxy M, Kapoor R, Terray P, Masson S (2014) Curious case of Indian Ocean warming. J Clim 27:8501–8509 https://doi.org/10.1175/JCLI-D-14-00471.s1

Roxy M, Ritika K, Terray P, Masson S (2015b) Indian Ocean warming: the bigger picture. Bull Am Meteorol Soc 96:1070–1071

Saji N, Goswami B, Vinayachandran PN, Yamagata T (1999) A dipole mode in the tropical Indian ocean. Nature 401:360–363

Schott, F. (1983). Monsoon response of the Somali Current and associated upwelling. *Progress in oceanography*, *12*(3), 357-381.

Schott FA, Xie SP, McCreary JP Jr (2009) Indian Ocean circulation and climate variability. Rev Geophys 47(1)

Shankar D, Vinayachandran P, Unnikrishnan AS (2002) The monsoon currents in the north Indian Ocean. Prog Oceanogr 52:63–120

Swapna P, Jyoti J, Krishnan R, et al (2017) Multidecadal Weakening of Indian Summer Monsoon Circulation Induces an Increasing Northern Indian Ocean Sea Level. Geophys Res Lett 44:10,560-10,572. https://doi.org/10.1002/2017GL074706

Swapna P, Krishnan R, Sandeep N, et al (2018) Long-Term Climate Simulations Using the IITM Earth System Model (IITM-ESMv2) With Focus on the South Asian Monsoon. J Adv Model Earth Syst 10:1127–1149. https://doi.org/10.1029/2017MS001262

Tourre Y, White WB (1995) ENSO Signals in global upper-ocean temperature. J Phys Ocean 25:1317–1332

Vinayachandran, P. N., Iizuka, S., & Yamagata, T. (2002). Indian Ocean dipole mode events in an ocean general circulation model. Deep Sea Research Part II: Topical Studies in Oceanography, 49, 1573-1596.

Wiggert JD, Hood RR, Banse K, Kindle JC (2005) Monsoon-driven biogeochemical processes in the Arabian Sea. Prog Oceanogr 65:176–213

Wijffels S, Meyers G (2004) An intersection of oceanic waveguides: variability in the Indonesian throughflow region. J Phys Oceanogr 34:1232–1253

Wyrtki K (1961) The thermohaline circulation in relation to the general circulation in the oceans. Deep Sea Res (1953) 8(1):39–64

Xie SP, Annamalai H, Schott FA, McCreary JP (2002) Structure and mechanisms of south Indian Ocean climate variability. J Clim 15:864–878

Yang J, Liu Q, Liu Z (2010) Linking observations of the Asian monsoon to the Indian Ocean SST: Possible roles of Indian Ocean Basin mode and dipole mode. J Clim 23:5889–5902. https://doi.org/10.1175/2010JCLI2962.1

Yang J, Liu Q, Xie SP, Liu Z, Wu L (2007) Impact of the Indian Ocean SST basin mode on the Asian summer monsoon. Geophys Res Lett 34:L02708 https://doi.org/10.1029/2006GL028571

Yu L, O'Brien JJ, Yang J (1991) On the remote forcing of the circulation in the Bay of Bengal. J Geophys Res Oceans 96:20449–20454. https://doi.org/10.1029/91jc02424

Zampieri M, Russo S, di Sabatino S, et al (2016) Global assessment of heat wave magnitudes from 1901 to 2010 and implications for the river discharge of the Alps. Science of the Total Environment 571:1330–1339. https://doi.org/10.1016/j.scitotenv.2016.07.008

Zhang L, Han W, Sienz F (2018) Unraveling causes for the changing behavior of the Tropical Indian Ocean in the past few decades. J Clim 31:2377–2388. https://doi.org/10.1175/JCLI-D-17-0445.1

Zheng J, Wang C (2019) Hot Summers in the Northern Hemisphere. Geophys Res Lett 46:10891–10900. https://doi.org/10.1029/2019GL084219

Zhou T, Yu R, Zhang J, et al (2009) Why the Western Pacific subtropical high has extended westward since the late 1970s. J Clim 22:2199–2215. https://doi.org/10.1175/2008JCLI2527.1

**Figures:**

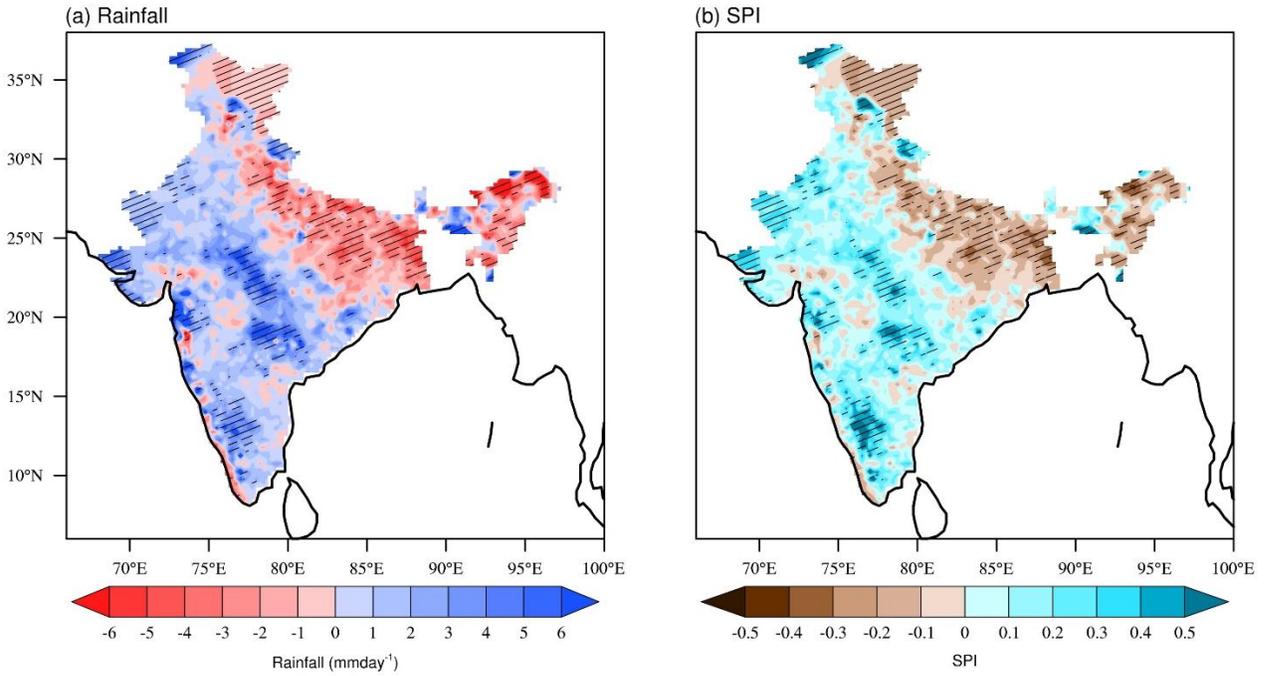

**Figure 1**: Anomalous (a) Rainfall (b) SPI index and the striped lines shows the value of significance at 90 % level computed using student t test in both the plots for the year 2022.

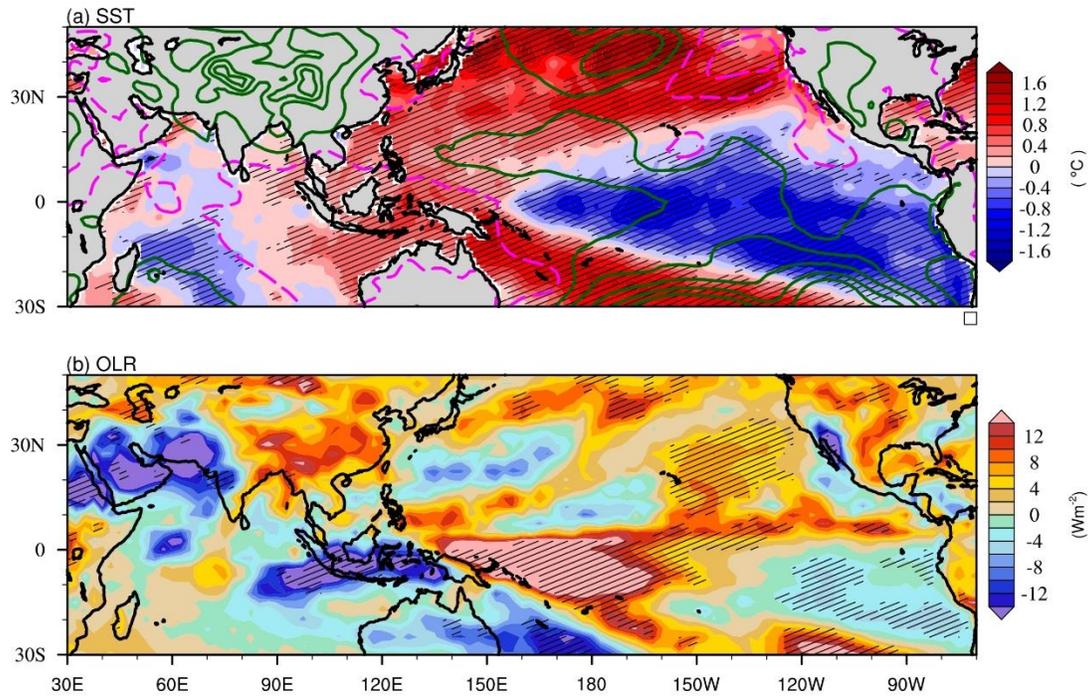

**Figure 2**: Anomalous value of (a)SST (shaded) and SLP (contour) (green contour are positive values and pink are negative value) (b)OLR. The striped lines show the value of significance at 90 % level computed using student t test for the year 2022.

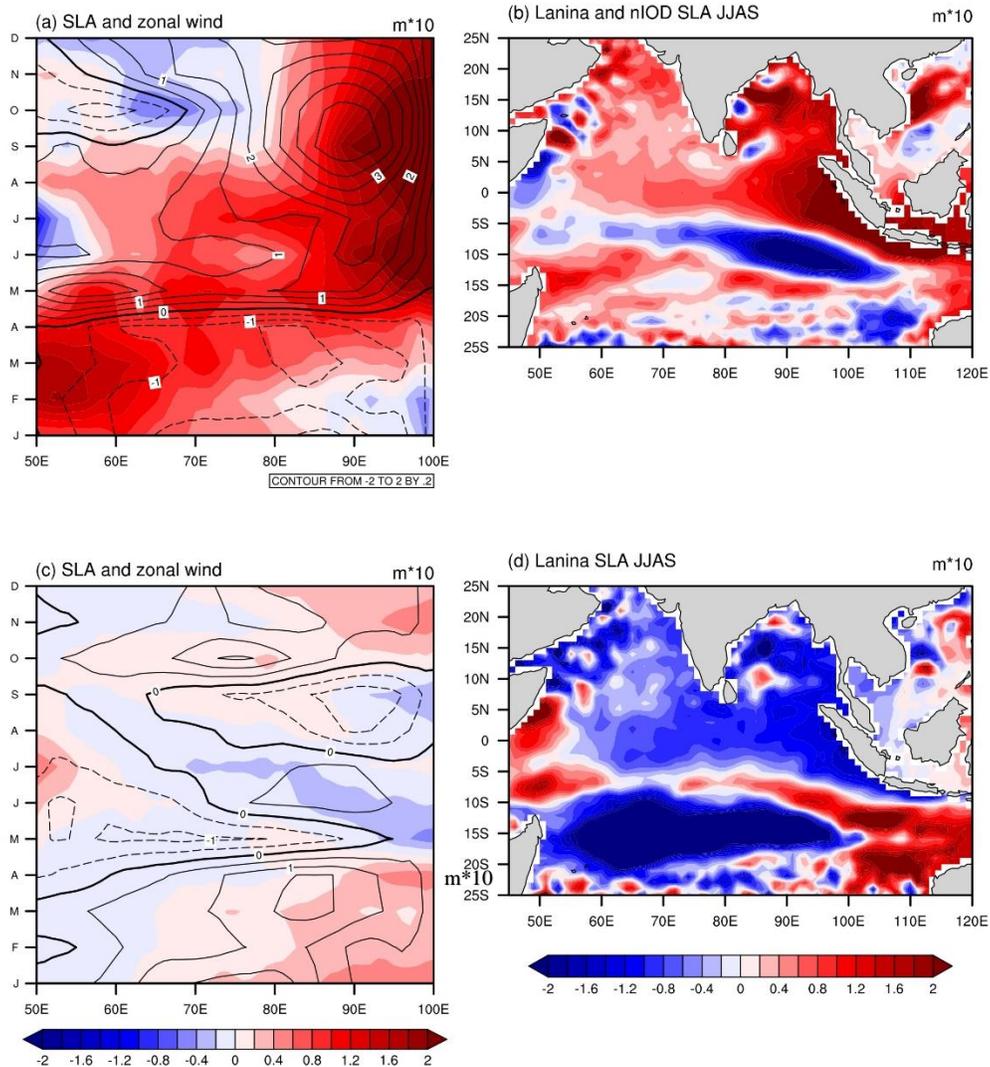

**Figure 3**: For co-occurrence events LaNina and nIOD (a) SLA JJAS (b) Hovmuller plot SLA shaded and zonal wind contour (solid lines are positive and dotted lines are negative contour) averaged over 2°N to 2°S. For pure LaNina events. (c) SLA JJAS (d) Hovmuller plot SLA shaded and zonal wind contour (solid lines positive and dotted lines negative) averaged over 2°N to 2°S.

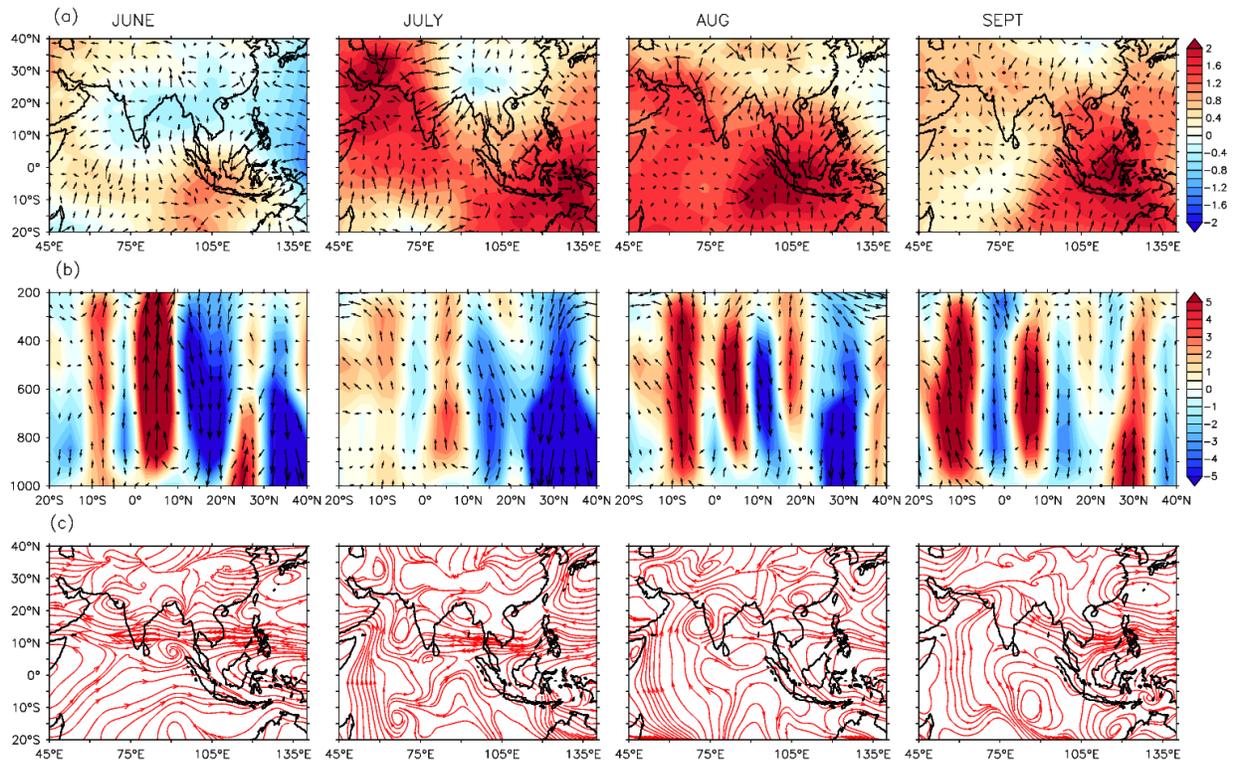

**Figure 4**: (a) Velocity potential with winds convergence and divergence. (b) The latitude-pressure section of monsoon Hadley-type circulation. The meridional and vertical velocities are averaged longitudinally between 80E and 100E. The shading denotes the magnitude of vertical velocity (-omega * 100) (c) Streamline at 850hPa for the year 2022.

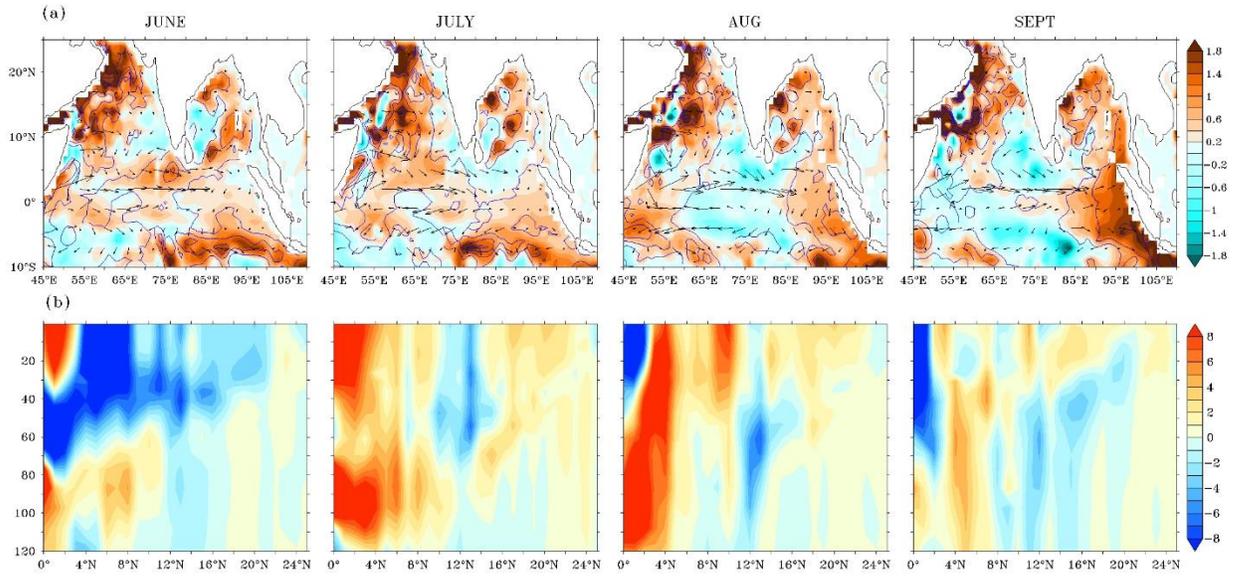

**Figure 5**: (a) Shows anomalous values of D20 (contours) and Ocean heat content (shaded) overlay with ocean currents. (b) Shows the meridional heat transport averaged over 45E to 75E for the year 2022

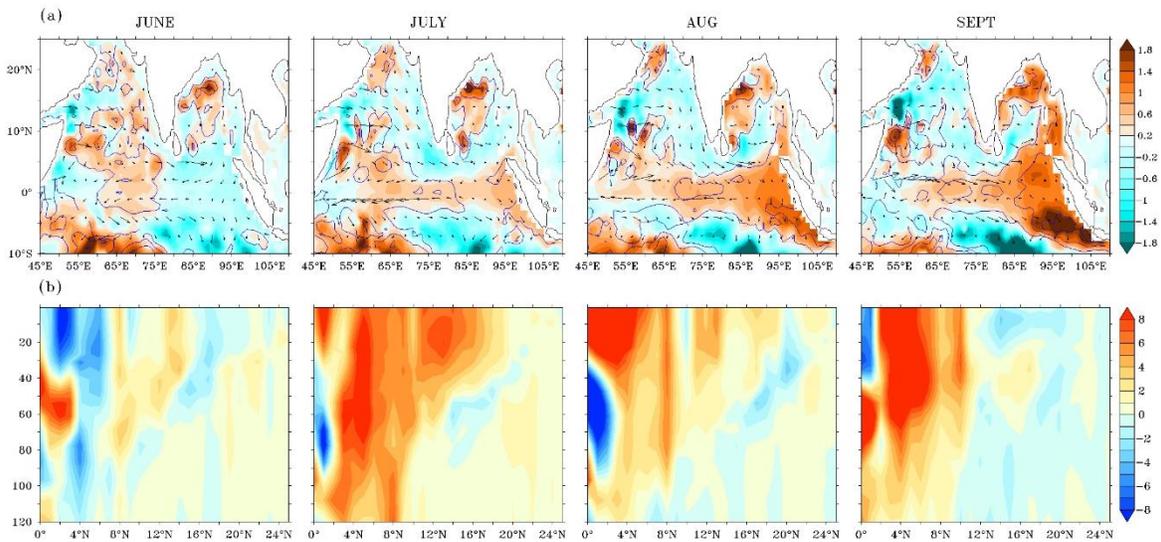

**Figure 6**: (a) Anomalous values of D20 (contours) and Ocean heat content (shaded) overlay with ocean currents. (b) The meridional heat transport averaged over 45E to 75E for the year 2010

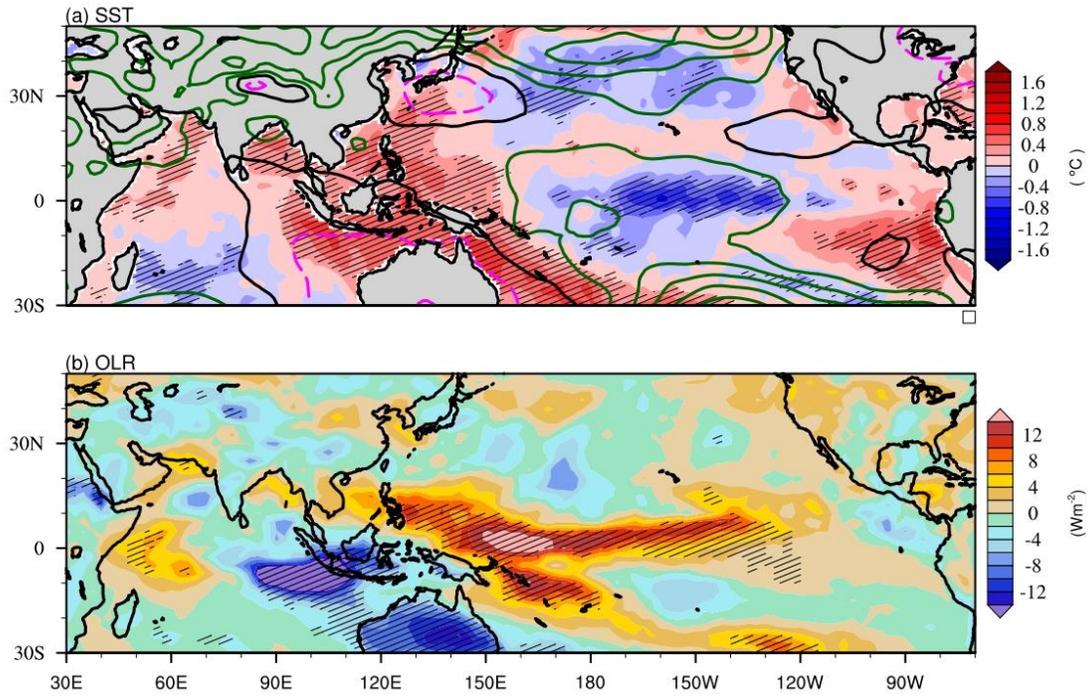

**Figure 7**: Anomalous value of (a)SST (shaded) and SLP (contour) (b)OLR. The striped lines show the value of significance at 90 % level computed using student t test for the observed composite years.

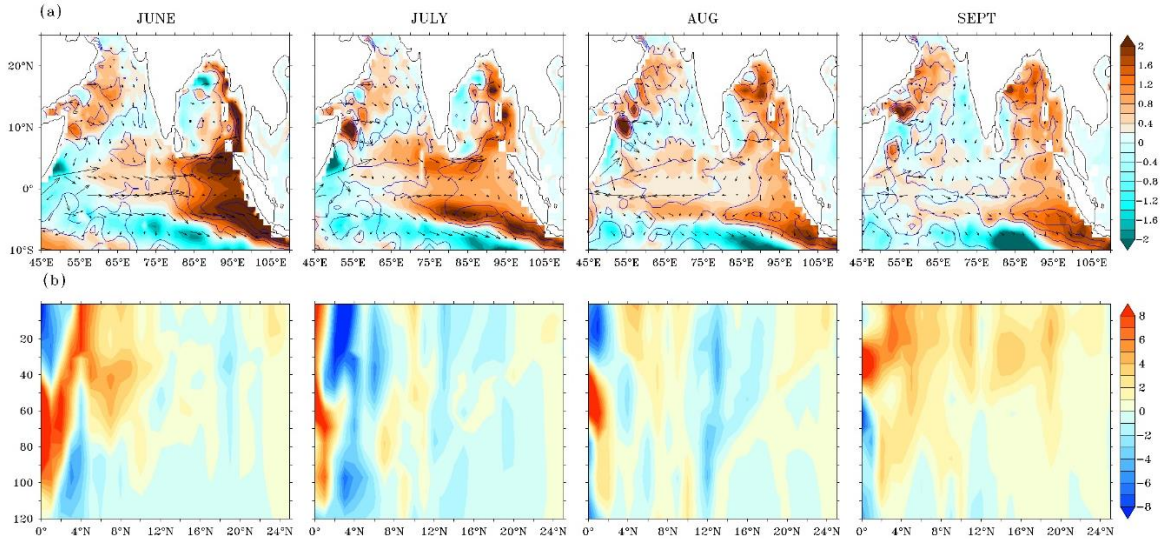

**Figure 8**: (a)Shows anomalous values of D20 (contours) and Ocean heat content (shaded) overlay with ocean currents. (b)Shows the meridional heat transport averaged over 45E to 75E for the observed composite years.

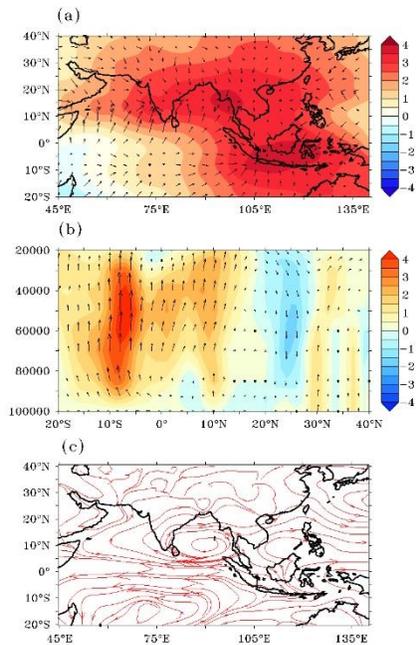

**Figure 9**: (a) Velocity potential with winds convergence and divergence. (b) The latitude-pressure section of monsoon Hadley-type circulation. The meridional and vertical velocities are averaged

longitudinally between 80E and 100E. The shading denotes the magnitude of vertical velocity (-omega * 100) (c) Streamline at 850hPa for the model composite years JJAS season.

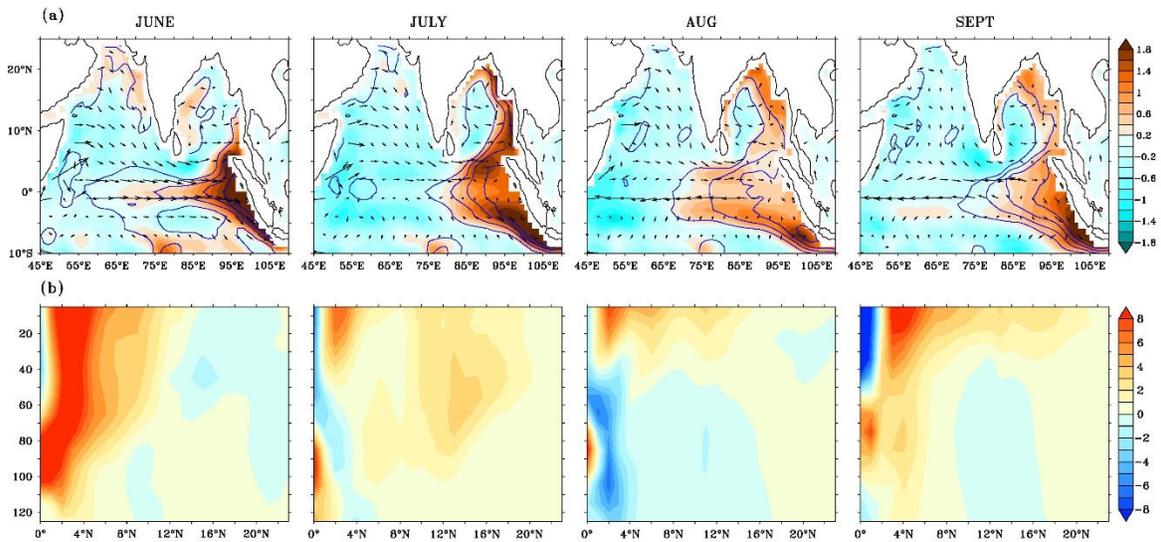

**Figure 10**: (a)Shows anomalous values of D20 (contours) and Ocean heat content (shaded) overlay with ocean currents. (b)Shows the meridional heat transport averaged over 45E to 75E for model composite years.

Supporting Information for

**Oceanic influence on Large-Scale Atmospheric Convection during co-occurring La Niña and IOD events**

Supriya Ovhal[1,2], Mujumdar M[1*], Swapna P[1], Sreenivas P[3], Sandeep N[1], M. Ravichandran[4]

[1] Centre for Climate Change Research, Indian Institute of Tropical Meteorology, Ministry of Earth Sciences (MoES), Pune 411008, India

[2] Department of Atmospheric and Space Sciences, Savitribai Phule Pune University, Pune, 411007, India

[3] University of Hyderabad, Telangana 500046, India

[4] Ministry of Earth Sciences (MoES), New Delhi, India

**Supplementary Figures**

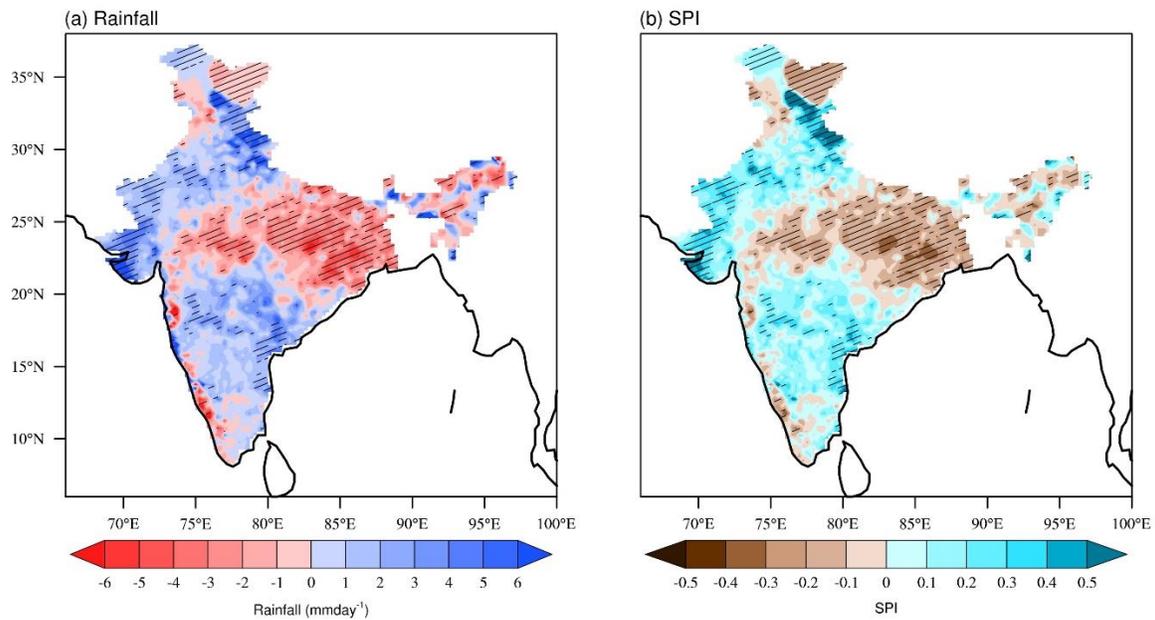

**Figure S1**: Anomalous (a) Rainfall (b) SPI index and the striped lines shows the value of significance at 90 % level computed using student t test in both the plots for the year 2010.

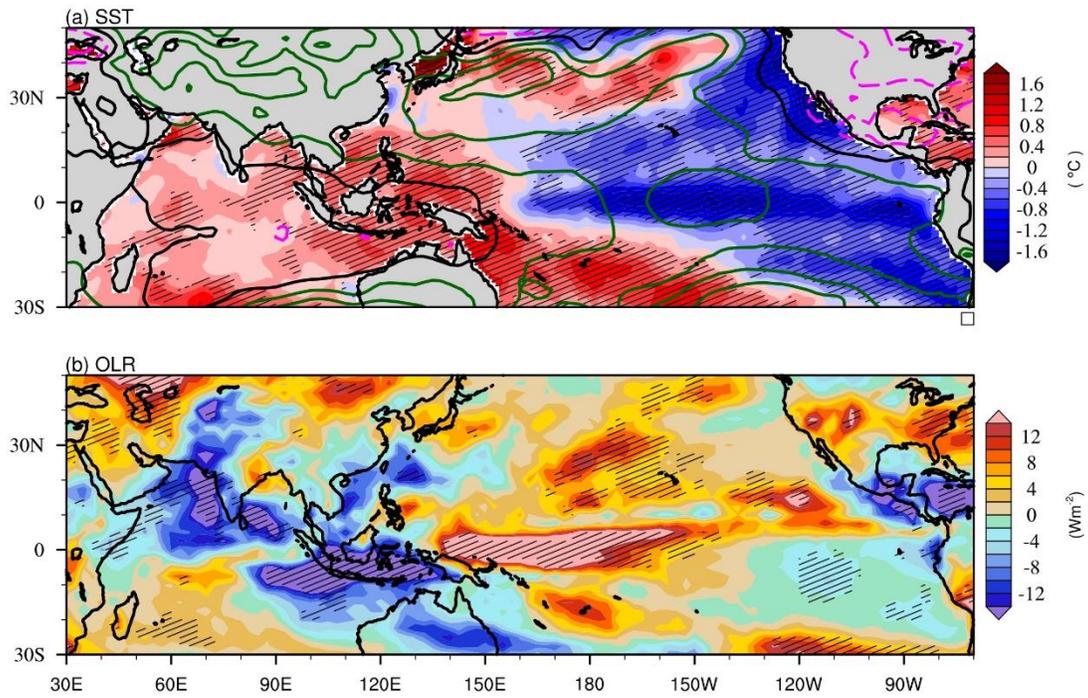

**Figure S2**: Anomalous value of (a)SST (shaded) and SLP (contour) (green contour are positive values and pink are negative value) (b)OLR. The striped lines show the value of significance at 90 % level computed using student t test for the year 2010.

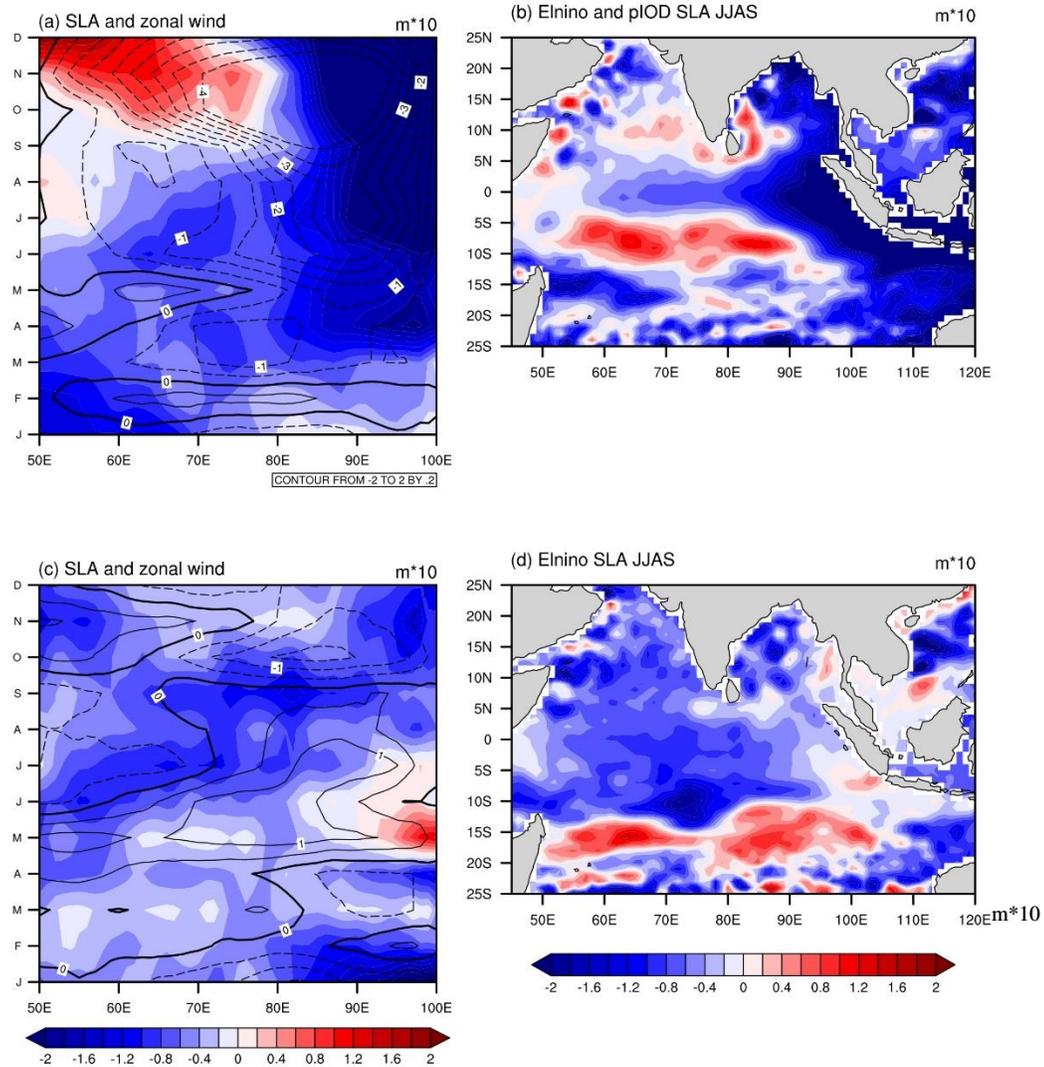

**Figure S3**: For co-occurrence events El Nino and pIOD (a) SLA JJAS (b) Hovmuller plot SLA shaded and zonal wind contour (solid lines are positive and dotted lines are negative contour) averaged over 2°N to 2°S. For pure El Nino events. (c) SLA JJAS (d) Hovmuller plot SLA shaded and zonal wind contour (solid lines positive and dotted lines negative) averaged over 2°N to 2°S.

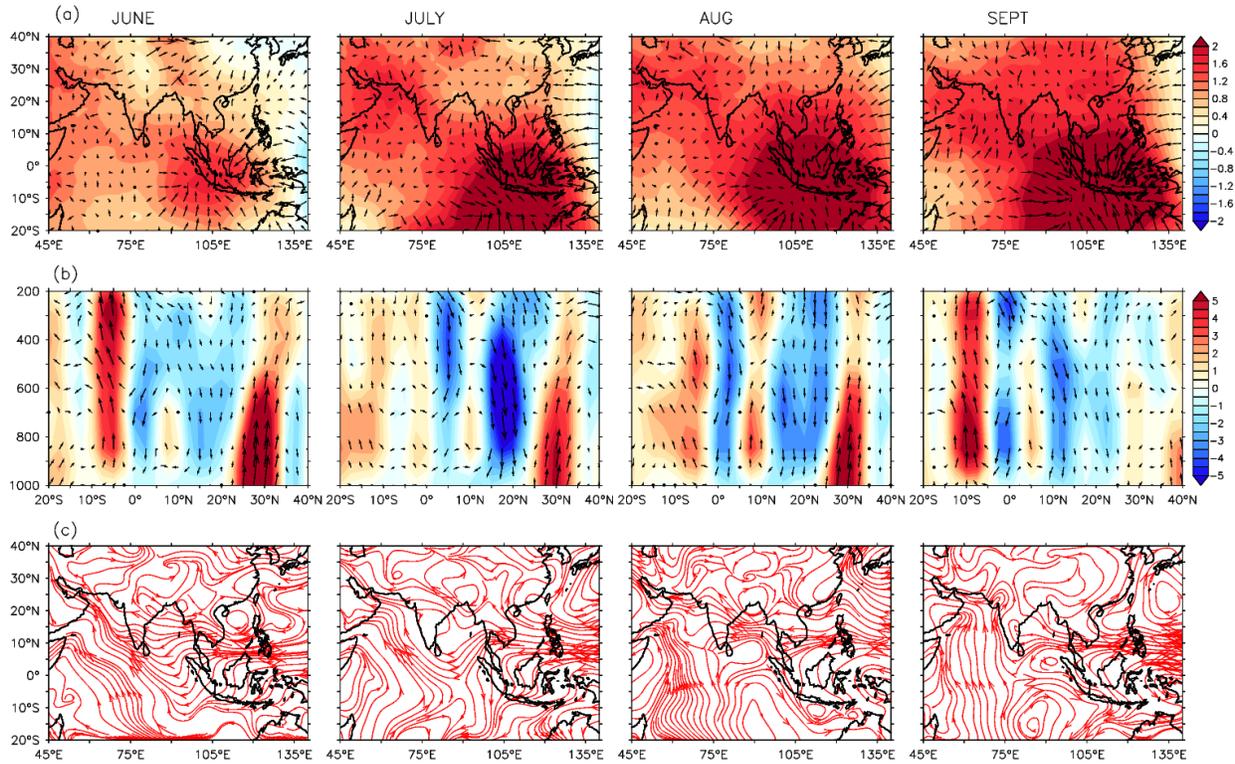

**Figure S4**: (a) Velocity potential with winds convergence and divergence. (b) The latitude-pressure section of monsoon Hadley-type circulation. The meridional and vertical velocities are averaged longitudinally between 80E and 100E. The shading denotes the magnitude of vertical velocity (-omega * 100) (c) Streamline at 850hPa for the year 2010.

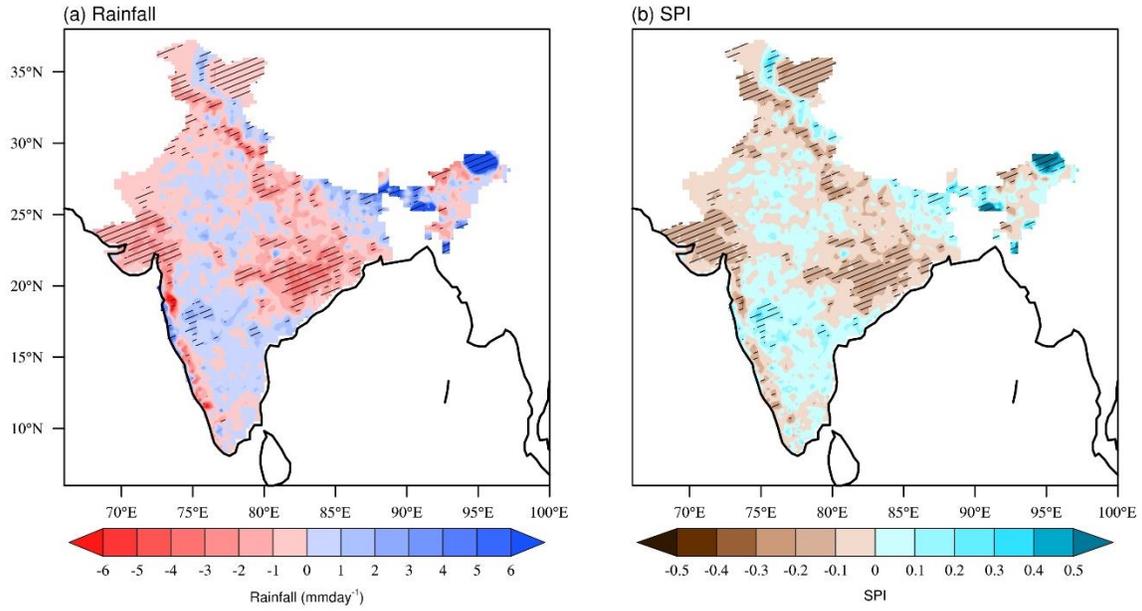

**Figure S5**: Anomalous (a) Rainfall (b) SPI index and the striped lines shows the value of significance at 90 % level computed using student t test in both the plots for the observed composite years.

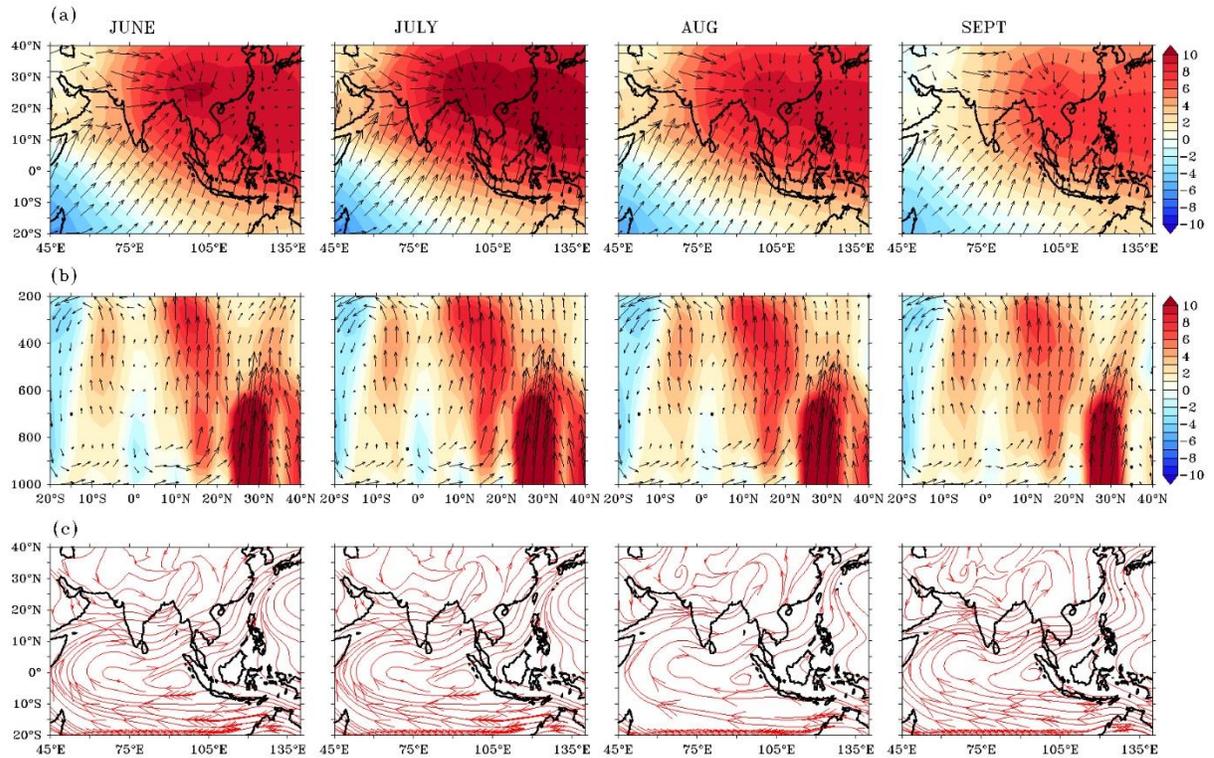

**Figure S6**: (a) Velocity potential with winds convergence and divergence. (b) The latitude-pressure section of monsoon Hadley-type circulation. The meridional and vertical velocities are averaged longitudinally between 80E and 100E. The shading denotes the magnitude of vertical velocity (-omega * 100) (c) Streamline at 850hPa climatology for the period 1974-2022.

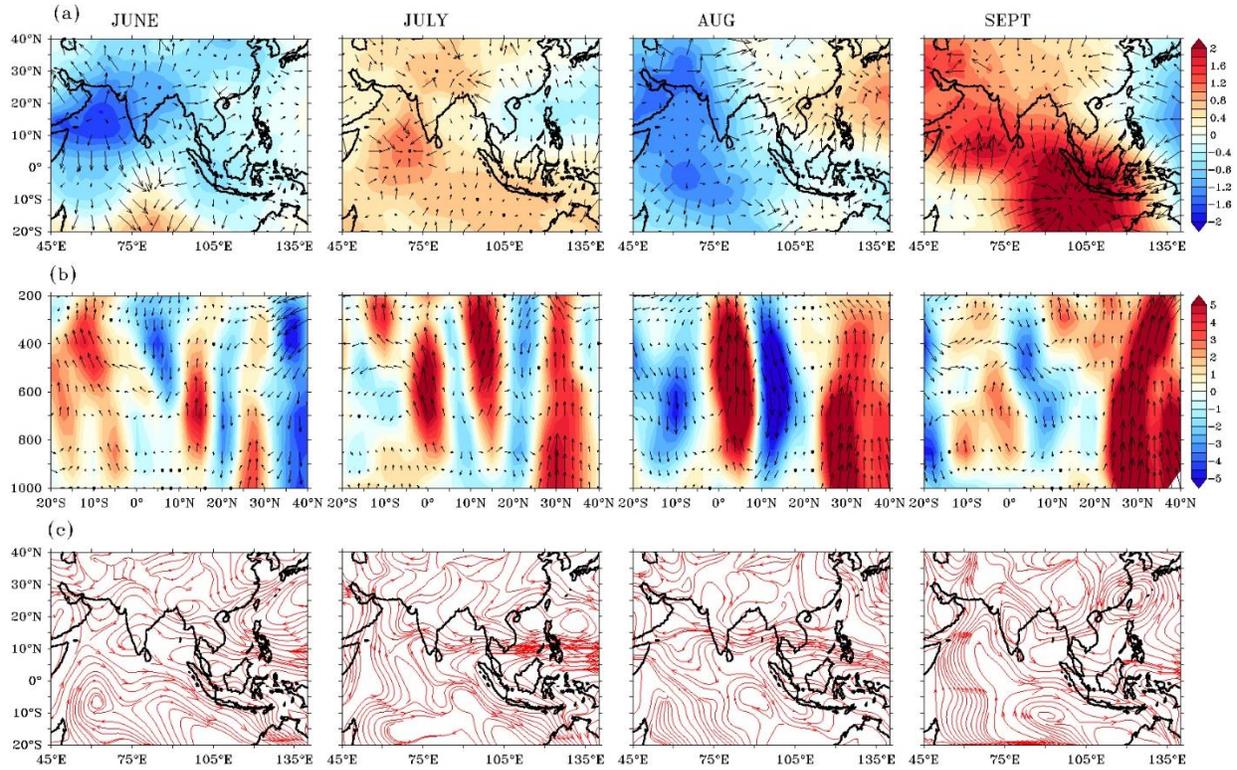

**Figure S7**:(a) Velocity potential with winds convergence and divergence. (b) The latitude-pressure section of monsoon Hadley-type circulation. The meridional and vertical velocities are averaged longitudinally between 80E and 100E. The shading denotes the magnitude of vertical velocity (-omega * 100) (c) Streamline at 850hPa for the observed composite years,

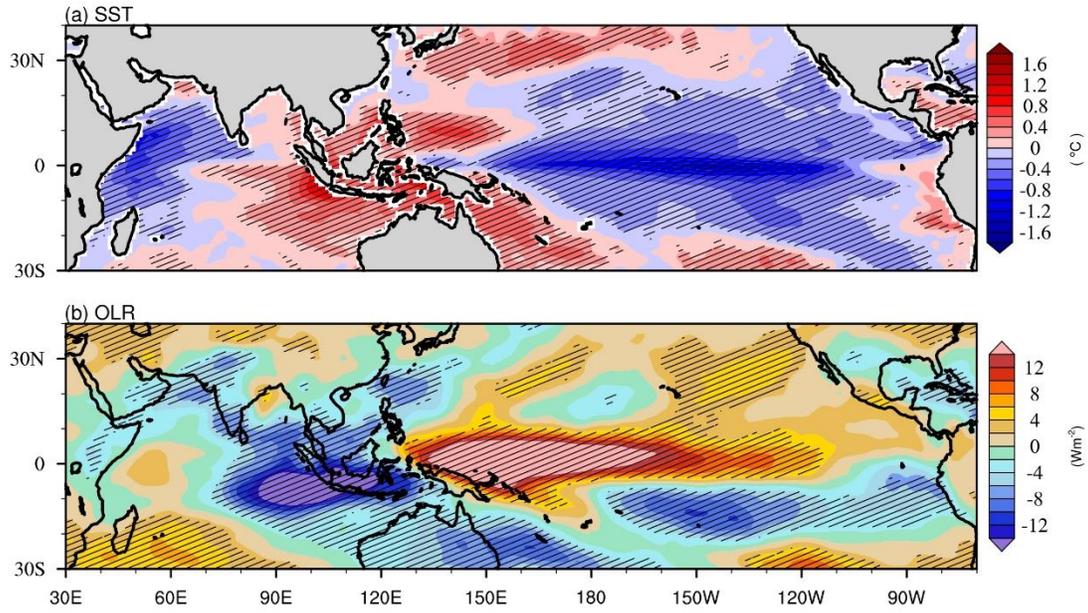

**Figure S8**: Anomalous value of (a)SST (shaded) and (b)OLR. The striped lines show the value of significance at 90 % level computed using student t test for the model composite years.